\documentclass[prb,twocolumn,showpacs,preprintnumbers,amsmath,amssymb,superscriptaddress]{revtex4}
\usepackage{graphicx}
\usepackage{dcolumn}
\usepackage{bm}

\newcommand{\BE}[1]{\begin{equation}\label{#1}}
\newcommand{\BEA}[1]{\begin{eqnarray}\label{#1}}
\newcommand{\BSE}[1]{\begin{subequations}\label{#1}}\let \= \equiv \let\*\cdot \let\~\widetilde \let\^\widehat \let\-\overline

\begin{document}
\preprint{APS/123-QED}
\title{Velocity Distributions of Tracer Particles in Thermal Counterflow in Superfluid $^4$He}
 \author{Y. Mineda}\affiliation{Department of Physics, Osaka City University, Sumiyoshi-ku, Osaka 558-8585, Japan}
\author{M. Tsubota}\affiliation{Department of Physics, The OCU Advanced Research Institute for Natural Science and Technology (OCARINA),
 Osaka City University, 3-3-138 Sugimoto, Sumiyoshi-ku, Osaka 558-8585, Japan}
\author{Y. A. Sergeev}\affiliation{School of Mechanical and Systems Engineering, University of Newcastle, Newcastle upon Tyne,
NE1 7RU, United Kingdom}
\author{C. F. Barenghi}\affiliation{School of Mathematics, University of Newcastle, Newcastle upon Tyne, NE1 7RU, United Kingdom}
\author{W. F. Vinen}\affiliation{School of Physics and Astronomy, University of Birmingham B15 2TT, United Kingdom}
\date{\today}
\begin{abstract}Quantum turbulence accompanying thermal counterflow in superfluid $^4$He was recently visualized by the Maryland group,  using  micron-sized  tracer particles of solid hydrogen (J. Phys. Soc. Jpn. {\bf 77}, 111007 (2008)) .  In order to understand the observations we formulate the coupled dynamics of fine particles and quantized  vortices,   in the presence of a relative motion of the normal and superfluid components. Numerical simulations based on this formulation  are shown to agree reasonably well  with experimental observations of the velocity distributions of the tracer particles in thermal counterflow.
\end{abstract}
\maketitle
\section{Introduction}
Quantum turbulence is a type of turbulence that is strongly affected by quantum effects.  It is observed in superfluids,  in which rotational flow of the superfluid component can take place only through the presence of quantized vortex filaments.  Turbulence in the superfluid component then takes the form of a tangle of these quantized vortices\cite{Hal,Tsu}. Quantum turbulence was first observed and studied\cite{Vin} in thermal counterflow in superfluid $^4$He.  Such counterflow is associated with two-fluid behaviour,  in which the superfluid component,  density $\rho_{\mathrm{s}}$, coexists with a normal component,  density $\rho_{\mathrm{n}}$, which carries all the entropy in the system {\cite{Lan}}.  The two components can have separate velocity fields, $\bm{v}_{\mathrm{s}}$ and $\bm{v}_{\mathrm{n}}$.  Heat is carried in such a system by counterflow of the two fluids, with no net mass flow,  the superfluid component moving towards the source of heat and the normal component away from it.  Above a certain critical heat current this counterflow proves to be unstable,  the superfluid component becoming turbulent. The normal fluid may also become turbulent,  but probably only at heat currents larger than those with which we are concerned in this paper.  This type of counterflow turbulence has no classical analogue.  The vortex lines present in the turbulent superfluid component interact with the normal fluid,  giving rise the force of "mutual friction" between the two fluids,  and study of the behaviour of this mutual friction provided the first experimental evidence for the existence of counterflow turbulence {\cite{Vin}}.  Since then,  many experimental,  theoretical and numerical studies of counterflow turbulence have been reported.  Schwarz carried out numerical simulations with the vortex filament model to confirm the existence of a self-sustained statistically steady turbulent state in the superfluid component{\cite{Sch2}}, and a more satisfactory simulation based on a full Biot-Savart simulation has recently been published by Adachi {\it et al.}{\cite{AFT}}. However, until recently there has been no direct visualization of the flow,  such as might convince a skeptic that thermal counterflow is a reality and might provide detailed information about the states of turbulent flow. We shall first describe briefly the two visualization techniques that have been developed recently,  and then focus on the interpretation of one particular set of experimental results.

The first technique relies on the production within the helium of micron-sized tracer particles formed from solid hydrogen,  and motion of these particles has been followed by both Particle Image Velocimetry {\cite{Zha}} and Particle Tracking Velocimetry {\cite{Bew,Pao}}.   The second technique {\cite{Guo}} uses metastable triplet state He$_2$ molecules as tracers.   These molecules are not trapped by vortex lines above 1K,  and they respond only to motion of the normal fluid;  they have been used recently to show that at a sufficiently large heat flux the normal fluid in thermal counterflow does become turbulent.  In this paper we shall be concerned with the interpretation of experiments on thermal counterflow that use as tracers particles of hydrogen;  these tracers can be trapped (intermittently) by vortex lines above 1K, so that interpretation is less straightforward than is the case for the molecules.  However,  the experiments by Paoletti \textit{et al.} \cite{Pao} with which we shall be concerned relate to heat fluxes that are rather small,  so that we are at least spared the need to consider the effect of a turbulent normal fluid.

In the experiments by Paoletti \textit{et al.} \cite{Pao} the tracer particles in a thermal counterflow were observed to divide into two groups: those that move freely in the direction of the normal fluid, and those that are trapped into vortex lines and move in the opposite direction. The velocity distribution of the trapped particles is broader, reflecting an irregular motion of the vortices. This difference was visible in the observed trajectories: smooth for the first group but more irregular for the second. The primary aim of this paper is to account for Paoletti's observations by modelling the coupled dynamics of particles and vortices and using the model in numerical simulations.  The motion of a free particle that is not trapped by vortices is determined by a Stokes drag from the viscous normal fluid and by inertial effects which arise from both superfluid and normal fluid. The motion of  a particle trapped by a vortex is determined by the Stokes drag and inertial effects, together with the effect of the tension in the vortex to which the particle is attached.    The motion of a vortex line depends on its shape and on the force of mutual friction with the normal fluid, but it can be modified by the presence of the particle. The problem of particle-vortex interaction can be tackled at different levels of approximation. At the most microscopic level\cite{Kiv,Sch,Fuji,Bar,San,Kiv1}, the close interaction of particle and vortex and the trapping of the particle onto the vortex require the calculation of a complex system of image vortices (mathematically, trapping is the reconnection of the vortex with its own image inside the particle).  Furthermore,  the particles of hydrogen are of irregular and unknown shape,  so that this kind of calculation is hardly relevant in practice.  As described in Section II,  we have therefore used a simple model which neglects the details of the particle-vortex interaction that occur on length scales of the order of the particle size or less, and which is sufficiently simple to allow the study at a useful and instructive level of the coupled dynamics of many particles and many vortices simultaneously.  In Section III we present an even simpler model,  applicable only at the smallest velocities,  when the configuration of vortex line is modified to a negligible extent by the presence of the particles.  We summarise our results in Section IV.

\section{Computations}

We use the vortex filament model{\cite{Sch1,AFT,Sch2}} and represent the vortex as strings of discrete points ${\bm s}$$\left( \xi, t \right)$, where $t$ is time and $\xi$ is arc length. The superfluid velocity at a point ${\bm r}$ due to a filament is given by the Biot-Savart expression:
\begin{equation}{\bm v}_{{\mathrm s}}\left( {\bm r}\right) = \frac{\kappa}{4\pi}\int_{\cal L}\frac{\left( {\bm s}_{1}-{\bm r}\right) \times d{\bm s}_{1}}{\left|  {\bm s}_{1}-{\bm r} \right|^3},
\label{eq:1a}
\end{equation}
where $\kappa$ is the quantum of circulation, ${\bm s}_{1}$ is a point on the filament, and integration is taken along the filament. Calculating the velocity ${\bm v}_{{\mathrm s}}$ at a point ${\bm r}={\bm s}$ on the filament causes the integral to diverge as ${\bm s}_{1}$ $\rightarrow$ ${\bm s}$. To avoid this divergence, we separate the velocity $\dot {\bm s}$ of the filament at the point ${\bm s}$ into two components{\cite{Sch1}}:
\begin{equation}\dot {\bm s} = \frac{\kappa}{4\pi}{\bm s}' \times {\bm s}'' \ln \left( \frac{2\left( l_{+}l_{-}\right)^{1/2}}{e^{1/2} \xi_{0}} \right)+\frac{\kappa}{4\pi} \int _{\cal L}' \frac{\left({\bm s}-{\bm r}\right) \times d{\bm s}}{\left| {\bm s}-{\bm r} \right|^{3}}.
 \label{eq:2a}
\end{equation}
The first term is the localized induction field arising from a curved line element acting on itself, and $l_{+}$ and $l_{-}$ are the lengths of the two adjacent line elements after discretization of $\Delta\xi$, separated by the point ${\bm s}$. The prime denotes differentiation with respect to the arc length $\xi$.  $\bm s'$, $\bm s''$, and $\bm s' \times \bm s''$ are the tangent vector, the normal vector, and binormal vector, respectively. The parameter $\xi_{0}$ is the cutoff corresponding to the core radius. The second term represents the non-local field obtained by integrating  Eq. (\ref{eq:1a}) along the rest of the filament, excluding the neighborhood of ${\bm s}$. We take account of the contributions from the boundary-induced velocity field ${\bm v}_{{\mathrm {s,b}}}$ and applied field ${\bm v}_{{\mathrm {s,a}}}$ in Eq. (\ref{eq:2a}).

\begin{figure}[h]
\begin{center}
\includegraphics[width=0.25\textwidth]{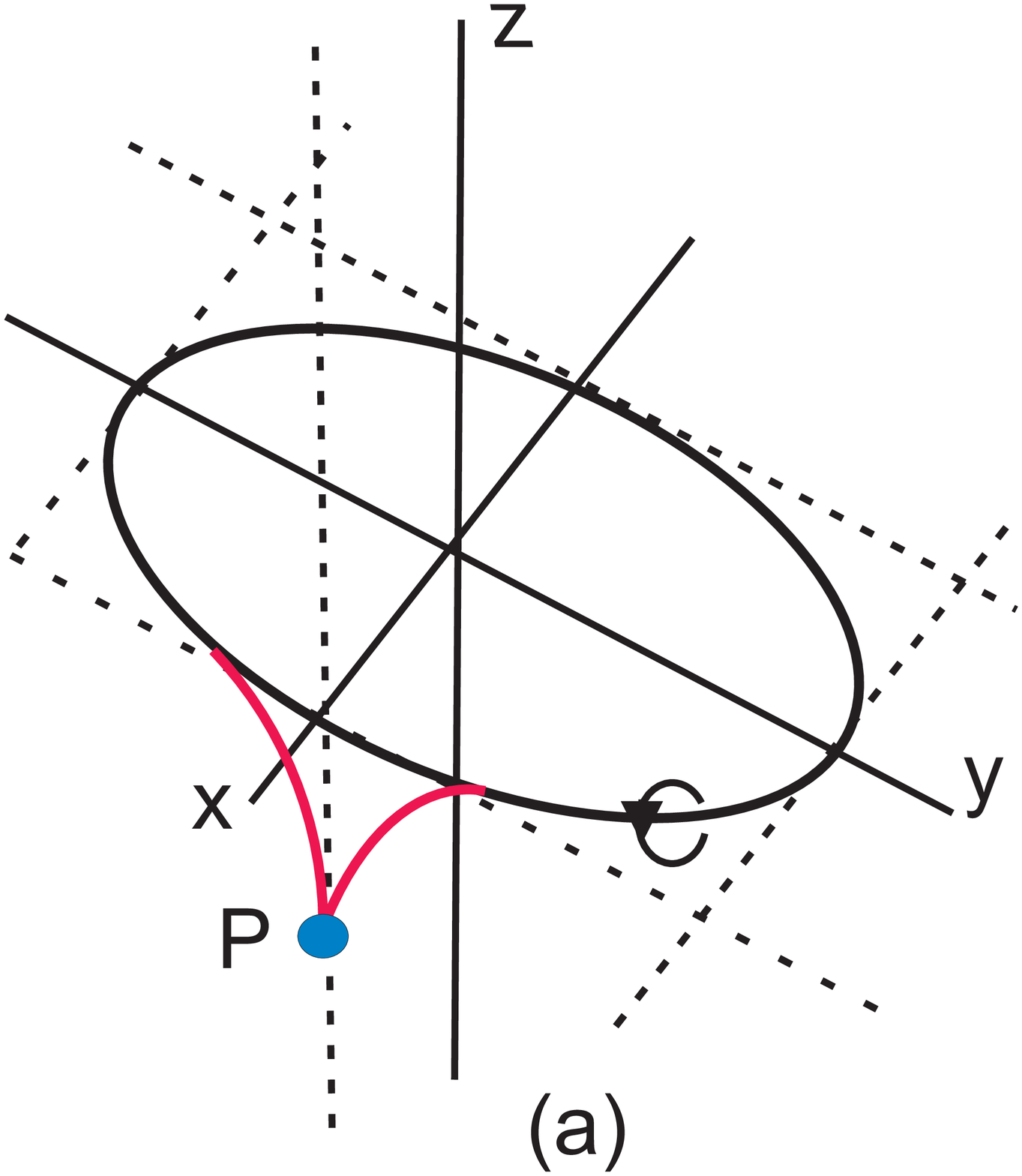}
\includegraphics[width=0.25\textwidth]{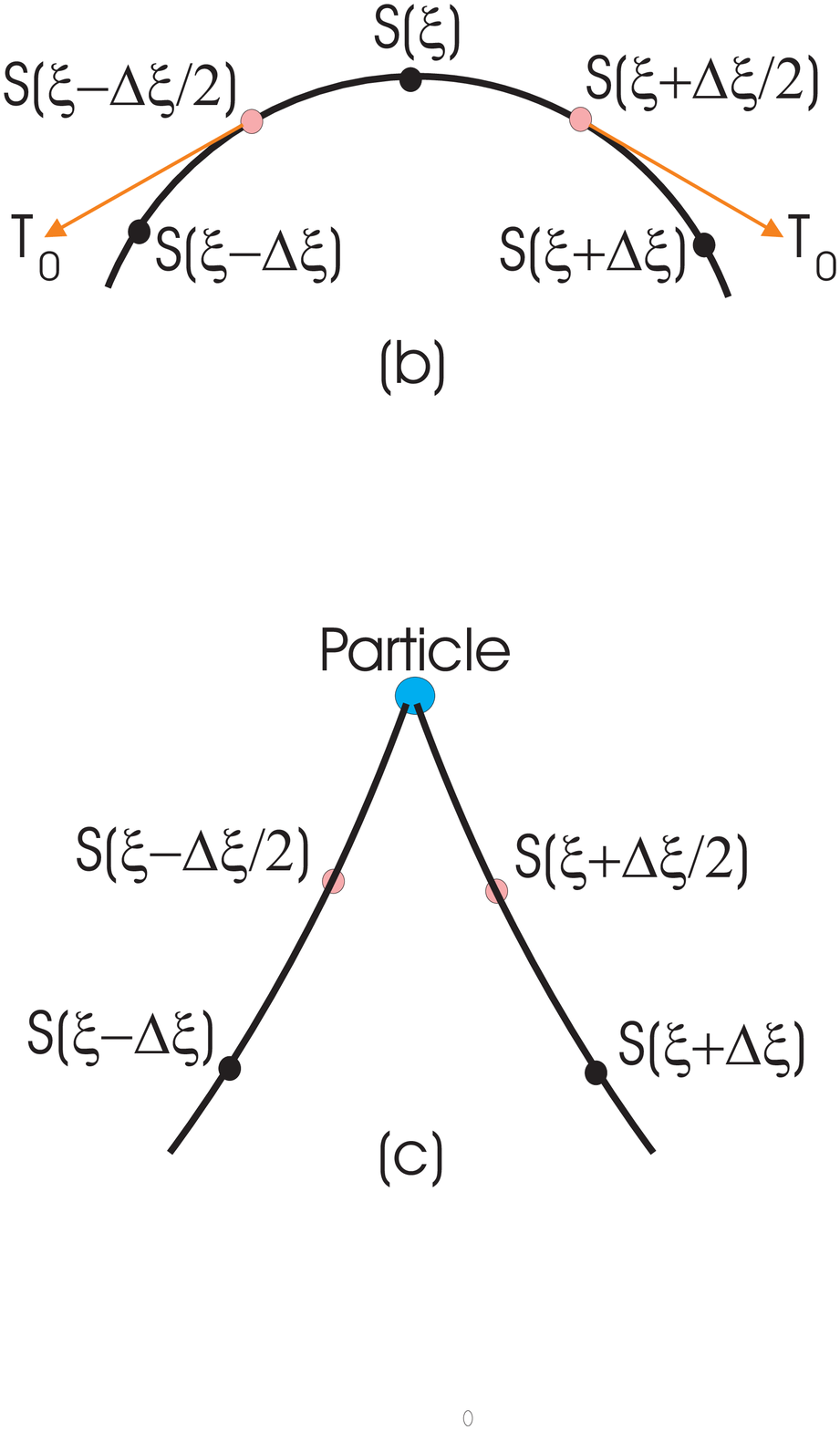}
\caption{(Color online) (a) Illustrating the way in which a vortex,  moving relative to the normal fluid, is distorted by a trapped particle;  a vortex ring,  instantaneously in the x-y plane,  propagates along the z-axis, with and without a trapped particle, the normal fluid being at rest ;  the ring gradually collapses,  under the combined influence of mutual friction and the viscous drag on the particle. (The (red) cusp to which the particle is attached points in the direction of $-z$.) (b) and (c) Schematic of the tension force acting on an element of vortex line,  with and without an attached particle.  We emphasize that (c) is schematic only;  the real vortex configuration in (c) requires a representation in three dimensions,  as in (a).}
\label{schem}
\end{center}
\end{figure}

We now explain how we handle to motion of a vortex line to which a particle is attached at a particular point.  The way in which a vortex line is distorted by a trapped particle is illustrated in Fig.1(a).   We shall assume that the vortex line is smooth on a scale of order the separation between the discrete points used in the computation,  except in the immediate neighbourhood of the particle,  and that the particle is situated at one of these points ${\bm s}(\xi)$  (Fig.1(c)).  The forces acting on the particle are:  a net force due to the tension in the sections of vortex line that adjoin the particle,  which is related to the angle between these two sections of line;  the drag force on the particle due to the viscosity of the normal fluid flowing relative to the particle; and inertial forces arising from both the superfluid component and the normal fluid;  and a Magnus force.  We note that the angle between the sections of vortex line adjoining the particle is not equal to the angle between the lines joining the particle point to the adjacent discrete vortex points, $s(\xi+\Delta\xi)$ and $s(\xi-\Delta\xi)$.  Furthermore,  we do not know the precise shape of the vortex in the immediate vicinity of the particle.  This potential difficulty can be handled in a number of different ways.  Here we shall handle it by considering the forces acting on the whole section of vortex line extending between the two mid points $s(\xi+\Delta\xi/2)$ and $s(\xi-\Delta\xi/2)$,  including the particle, regarded as a single entity.

To make our procedure more clear we shall first consider the forces acting on this section of vortex line in the absence of the particle; in this case there can be no kink in the vortex line at $s(\xi)$ (Fig.1(b)).  First there is a force, $F_T$ due to the tension in the line,  given by
\begin{equation} {\bm F}_{\mathrm T} =  T_0[{\bm s'}(\xi+\Delta\xi/2)-{\bm s'}(\xi-\Delta\xi/2)],
\label{eq:3a}
\end{equation}
 where $T_0$ is the line tension,  given by
\begin{equation} T_{0} = \frac{\kappa}{4\pi}\ln \left( \frac{2\left( l_{+}l_{-}\right)^{1/2}}{e^{1/2} \xi_{0}} \right).
\label{eq:4a}
\end{equation}
In a computation we make the approximation that $s'(\xi + \Delta\xi/2)$ is equal to the slope of the straight line joining the points $s(\xi)$ and $s(\xi + \Delta \xi)$.  Secondly there is a Magnus force,  given by
\begin{equation}  {\bm F}_{\mathrm M} = \rho_s\kappa {\bm s'}\times[\dot{\bm{s}}(\xi)-{\bm v}_{\mathrm{{\bm s},nl}}]\Delta\xi,
\label{eq:5a}
\end{equation}
where ${\bm v}_{\mathrm{s,nl}}$ is the second term in Eq.(\ref{eq:2a}) evaluated at the point ${\bm s}(\xi)$.  Since the mass of the core of a vortex is negligibly small,  these two forces must balance,  provided that there is no mutual friction.  Therefore the velocity $\dot{{\bm s}}$ is given by
\begin{equation}  {\bm F}_{\mathrm T} + {\bm F}_{\mathrm M} =0.
\label{eq:6a}
\end{equation}
If the shape of the vortex is sufficiently smooth,  and if $\Delta\xi$ is sufficiently small,  the first term in Eq.(\ref{eq:6a}) is equal to $T_0 {\bm s}''\Delta\xi$,  so that Eq.(\ref{eq:6a}) yields a velocity $\dot{\bm s}(\xi)$ similar to that given by Eq.(\ref{eq:2a}),  but derived in a different way.  The only difference is that  Eq.(\ref{eq:6a}),  in contrast to Eq.(\ref{eq:2a}),  does not  yield any contribution from the non-local field to the component of $\dot{{\bm s}}$ in a direction along the vortex line;  this component does not have any real physical effect on the vortex dynamics, although it can lead to a motion of the discrete vortex points.

Our analysis so far has assumed that the temperature is vanishingly small.  At a finite temperature a force of mutual friction,  given by
\begin{equation}
 {\bm F}_{\mathrm{MF}}= \Big(-\gamma_0 {\bm s}'\times[{\bm s}'\times({\bm v}_{\mathrm n} - \dot{{\bm s}})] -    \gamma'_0 {\bm s}'\times({\bm v}_{\mathrm n} - \dot{{\bm s}})\Big)\Delta\xi,
\label{eq:7a}
\end{equation}
must be added to the balance of forces,  where $\gamma_0$ and $\gamma'_0$ are temperature-dependent parameters.  Therefore $\dot{s}$ is now given by
\begin{equation}       {\bm F}_{\mathrm T} + {\bm F}_{\mathrm M} + {\bm F}_{\mathrm{MF}} = 0.
\label{eq:8a}
\end{equation}
It is easily shown that $\dot{s}$ is then given by
\begin{equation}   \dot{{\bm s}} = \dot{{\bm s}}_0 +\alpha {\bm s'} \times ({\bm v}_{\mathrm{n}} -  \dot{{\bm s}}_0) - \alpha' {\bm s'} \times [{\bm s'} \times ({\bm v}_{\mathrm{n}} -  \dot{{\bm s}}_0)],
\label{eq:9a}
\end{equation}
where $\dot{{\bm s}}_0$ is given by Eq.(\ref{eq:6a}), and $\alpha$ and $\alpha'$ are the more usual mutual-friction parameters.

Now consider the behaviour of a spherical tracer particle, radius $a_{\mathrm{p}}$ ($ < 0.5 \Delta\xi$) and volume $V_{\mathrm{p}}$.   The density of the material of the particle is assumed to be the same as that of the helium,  so that the particle is neutrally buoyant.  Consider first the motion of the particle when it is not trapped on a vortex.  Its equation of motion is
\begin{equation}  \frac{3}{2} \rho V_{\mathrm{p}} \frac{d{\bm v}_{\mathrm{p}}}{dt}={\bm f}_{\mathrm{S}} + {\bm f}_{\mathrm{I}}.
\label{eq:10a}
\end{equation}
We have allowed for the fact that the effective mass of the particle is enhanced in the fluid by an amount equal to half the mass of fluid displaced.  The force ${\bm f}_{\mathrm{S}}$ is the Stokes drag associated with the viscosity of the normal fluid,  given by
\begin{equation}  {\bm f}_{\mathrm{S}} = 6\pi a_{\mathrm{p}} \mu_{\mathrm{n}}({\bm v}_{\mathrm{n}} - {\bm v}_{\mathrm{p}}),
\label{eq:11a}
\end{equation}
where $\mu_{\mathrm{n}}$ is the viscosity of the normal fluid.  We have assumed that the particle Reynolds number is small,  and that the mean free path of the excitations constituting the normal fluid is much smaller than $a_{\mathrm p}$. The force ${\bm f}_{\mathrm{I}}$ is an inertial force acting on the particle associated with any acceleration of the normal and superfluid components,  and it is given by
\begin{equation}  {\bm f}_{\mathrm I} = \frac{3}{2}\rho_{{\mathrm s}}V_{{\mathrm p}}\frac{D {\bm v}_{{\mathrm s}}}{Dt}+\frac{3}{2}\rho_{{\mathrm n}}V_{{\mathrm p}}\frac{D {\bm v}_{{\mathrm n}}}{Dt}.
\label{eq:12a}
\end{equation}
 In our simulations we neglect the inertial force on the normal fluid,  because $v_n$ is assumed to be constant in time and spatially uniform.

Finally we consider the situation when the tracer particle is trapped on a vortex at the point ${\bm s}(\xi)$ (Fig.1(c)).  The element of vortex between the points ${\bm s}(\xi+\Delta\xi/2)$ and ${\bm s}(\xi-\Delta\xi/2)$ is now subject to the extra forces $f_{\mathrm{S}}$ and $f_{\mathrm{I}}$,  and the total force on this element must lead to the particle acceleration $d{\bm v}_{\mathrm p}/dt$;  i.e.
\begin{equation}  \frac{3}{2} \rho V_{\mathrm p} \frac{d{\bm v}_{\mathrm p}}{dt} = {\bm F}_{\mathrm T} + {\bm F}_{\mathrm M} + {\bm F}_{\mathrm MF} + {\bm f}_{\mathrm S} + {\bm f}_{\mathrm I}.
\label{eq:13a}
\end{equation}
It is important to understand that in calculating the forces ${\bm F}_{\mathrm T}$,  ${\bm F}_{\mathrm M}$  and  ${\bm F}_{\mathrm MF}$ we must take the element of vortex between the points ${\bm s}(\xi+\Delta\xi/2)$ and ${\bm s}(\xi-\Delta\xi/2)$ to have the shape that it has when the particle is attached;  this shape is not known in detail,  and it must involve,  for example,  a discontinuity in the slope ${\bm s}'$ at the position of the particle.  To make further progress we must therefore make approximations.  We shall make two assumptions,  which cannot be accurately correct;  we can hope that they do not  introduce serious errors,  although,  as we shall see,  there is some evidence that they do lead to small errors.  First,  we shall assume that the element of vortex moves with an average velocity equal to that of the particle,  ${\bf v}_{\mathrm p}$.  And secondly we shall assume that the effective length of the element,   relevant to the calculation of ${\bf F}_{\mathrm M}$ and  ${\bf F}_{\mathrm MF}$,  is not significantly different from that shown by the smooth curve in Fig.1(c).  We judge that these assumptions are likely to be good if the particle is significantly smaller than $\Delta\xi$.  Then the values of ${\bf F}_{\mathrm M}$ and  ${\bf F}_{\mathrm MF}$ are given by Eqs.(\ref{eq:5a}) and (\ref{eq:7a}),  but with $\dot{{\bf s}}$ replaced by ${\bf v}_{\mathrm p}$,  while ${\bf F}_{\mathrm T}$ is still given by Eq.(\ref{eq:3a}).   Therefore
\begin{eqnarray}
&&\frac{3}{2} \rho V_{\mathrm{p}} \frac{d{\bm v}_{\mathrm p}}{dt} =  6\pi a_{\mathrm p} \mu_{\mathrm n} ({\bf v}_{\mathrm n} - {\bf v}_{\mathrm p}) +  \frac{3}{2} \rho_{\mathrm s} V_{\mathrm p} \frac{D{\bm v}_{\mathrm s}}{Dt} +{} \hspace{15mm}   \nonumber  \\
&&{}  T_0[{\bm s'}(\xi+\Delta\xi/2)-{\bm s'}(\xi-\Delta\xi/2)] + \nonumber \\
&&{} \rho_{\mathrm s} \kappa {\bm s}' \times ({\bm v}_{\mathrm{p}} - {\bf v}_{\mathrm {s,nl}})\Delta \xi +{} \nonumber \\
&&{} \Big(\gamma_{0} {\bm s}'\times[{\bm s}'\times({\bm v}_{\mathrm p}-{\bf v}_{\mathrm n})] + \gamma'_{0} {\bm s}'\times ({\bm v}_{\mathrm p}-{\bf v}_{\mathrm n})\Big) \Delta\xi
\label{eq:14a}
\end{eqnarray}

We make the important remark that the acceleration given by Eq.(\ref{eq:14a}) ought not to depend on the magnitude of $\Delta \xi$.  A reduction in $\Delta\xi$ reduces the magnitudes of the last two terms on the right hand side of this equation (the Magnus force and the force of mutual friction),  both of which are proportional to $\Delta\xi$,  but this reduction is cancelled by an increase in tension term $T_0[{\bm s'}(\xi+\Delta\xi/2)-{\bm s'}(\xi-\Delta\xi/2)]$.

We note that,  correctly,  the component of  ${\bm v}_{\mathrm{s,nl}}$ parallel to the vortex makes no contribution to the velocity of the particle, although, as we have already noted,  it does make a contribution to the motion of the discrete vortex points that are not attached to a particle.

In summary,  then,  the dynamics of the untrapped particles is described by Eq.(\ref{eq:10a}),  and coupled dynamics of the vortices and trapped particle is described by Eq.(\ref{eq:14a}),  where ${\bf v}_{\mathrm {s,nl}}$ is given by the second term in Eq.(\ref{eq:2a}).

\begin{figure} [htbp]
\begin{center}
\includegraphics[width=0.9\linewidth]{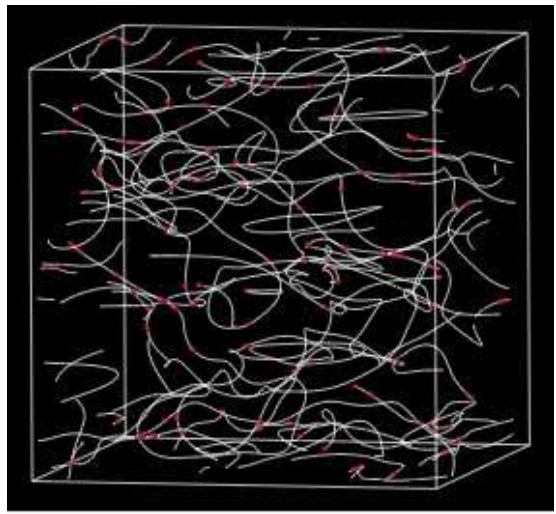}
\caption{(Color online) A snapshot of the coupled dynamics of the tracer particles and the vortices at $T$=1.9 K.   $\Delta\xi = 2.0\times 10^{-4}$ cm.  All of the 120 particles are trapped by vortices. The direction of the normal flow ${\bm v}_{{\mathrm n}}=0.30$ cm/s is upward, and that of superflow ${\bm v}_{{\mathrm s}}$ given by Eq. ({\ref{eq:15}}) is downward. The vortex line density is $L \sim 6000$ cm$^{-2}$.}  \label{fig:1}
\end{center}
\end{figure}

In practice,  the terms involving accelerations in Eq.(\ref{eq:14a}) are small and can be neglected,  so that the size of the particle enters only  in the Stokes drag as $a_{\mathrm{p}}$.   In our simulations we have taken as $a_{\mathrm{p}}=1.0\times 10^{-4}$ cm.  We have used two values of the numerical space resolution $\Delta\xi$:  $2.0\times 10^{-4}$ cm and $8.0\times 10^{-4}$ cm.  Our time resolution is $\Delta t = 1.0\times 10^{-6}$ s.  We use periodic boundary conditions with a box of size $0.1^3$ cm$^3$.  We have used three temperatures:  $1.9$ K,  $2.0$ K, and $2.1$ K.  The velocity of the normal fluid is assumed constant in time and spatially uniform. First we make thermal counterflow quantum turbulence that is statistically steady and satisfies the well-known relation $L = \gamma ^2v_{\mathrm {ns}}^2$ where $L$ is the vortex line density and $\gamma$ is a temperature-dependent parameter\cite{AFT}.  After that, we place 120 particles trapped on the vortices (Fig. \ref{fig:1}).  The turbulence continues to evolve in a steady state,  and we study the distribution of particle velocities.  A particle velocity is obtained from the distance moved by the particle in a fixed time interval, taken to be 10 ms when  $\Delta\xi =2.0\times 10^{-4}$ cm and 100 ms when  $\Delta\xi = 8.0\times 10^{-4}$ cm.  Some particles become detrapped,  and these are ignored;  this detrapping will be considered in a later paper.

We note that our values of $\Delta\xi$ are not large compared with $a_{\mathrm{p}}$,  contrary to an assumption made in our derivation of  Eq.(\ref{eq:14a}).  These small values were chosen in order to ensure as far as we can that our simulations describe correctly the evolution of sharp bends in the vortices.  Strictly speaking,  therefore,  our simulations relate to particles that are significantly smaller than $10^{-4}$ cm in radius,  but which experience a Stokes drag equal to that experienced by the larger particle.  Although smaller particles are likely to be detrapped more easily,  we can hope that the behaviour of the trapped particles is not significantly affected.

\begin{figure} [htbp]
\begin{center}
\includegraphics[width=1.0\linewidth]{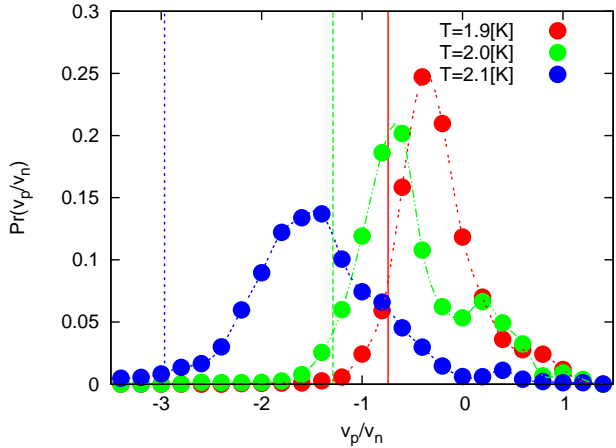}
\caption{(Color online) Velocity distributions of the particles in the direction of  the flow of the the normal fluid.  $\Delta\xi = 2.0 \times 10^{-4}$ cm.  The velocity ${\bm v}_{\mathrm p}$ is normalized by the normal fluid velocity ${\bm v}_{\mathrm n}$. The red, green and blue points and lines are at $T$=1.9 K, 2.0 K, 2.1 K, respectively,  and the corresponding normal fluid velocities are $0.30$,  $0.20$ and $0.09$ cm s$^{-1}$, respectively (corresponding to approximately the same vortex line density).  The vertical lines refer to the velocity of the superfluid given by Eq. (\ref{eq:15}).}
\label{fig:2}
\end{center}
\end{figure}

In Fig,\ref{fig:2} we show examples of the components of the velocity distributions of the tracer particles parallel to velocity of the normal fluid.  Superfluid velocities are calculated from the equation
\begin{equation}{\bm v}_{{\mathrm s}} = -\frac{\rho_{\mathrm n}}{\rho_{\mathrm s}}{\bm v}_{n}.
\label{eq:15}
\end{equation}

We see that the velocity of the particles is significantly slower than that of the vortices,  which move at a velocity that is only slightly smaller than that of the superfluid.  This feature is probably associated with a sliding of the particles along the vortices,  as we shall see more clearly in the next Section.

\section{Behaviour of the trapped particles in the limit of small velocities}

When velocities are sufficiently small,  the Stokes drag on a particle is insufficient to cause significant distortion of the vortices.  In that case the velocity with which a particle moves can be calculated by assuming that it simply slides along the vortex to which it is attached,  the properties of the vortex array being determined from simulations carried out in the absence of particles.

\begin{figure}[h]
\begin{center}
\includegraphics[width=0.6\linewidth]{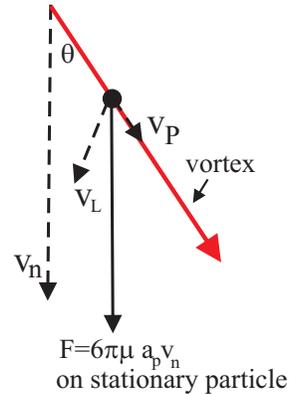}
\caption{(Color on line) Illustrating the sliding of a particle along a vortex.}
\label{sliding}
\end{center}
\end{figure}

Consider the behaviour of one particle under these conditions on a vortex line that is inclined at an angle $\theta$ to the direction of flow of the normal fluid (Fig.\ref{sliding}).  We need first to calculate the velocity $v_{\mathrm{p}}$ of the particle along the vortex,  when the vortex is itself moving with the velocity $v_{\mathrm{L}}$,  given by Eq.(\ref{eq:2a}),  evaluated in the absence of any trapped particles;  $v_{\mathrm{L}}$ is not necessarily in the plane defined by the vectors  $v_{\mathrm{p}}$ and $v_{\mathrm{n}}$. The Stokes force acting on the particle is given by the vector equation
\begin{equation}  \mathbf{F} = 6\pi\mu_{\mathrm{n}}a_{\mathrm{p}}(\mathbf{v}_{\mathrm{n}} - \mathbf{v}_{\mathrm{p}} - \mathbf{v}_{\mathrm{L}\bot}),
\label{eq:S1}
\end{equation}
where  $\mathbf{v}_{\mathrm{L}\bot}$ is the component of $\mathbf{v}_{\mathrm{L}}$ perpendicular to the vortex line.  The component of this force along the vortex line is
\begin{equation}  F_{\mathrm{L}} = 6\pi\mu_{\mathrm{n}}a_{\mathrm{p}}(v_{\mathrm{n}}\cos \theta - v_{\mathrm{p}}).
\label{eq:S2}
\end{equation}
This component must vanish.  Therefore
\begin{equation}  v_{\mathrm{p}} = v_{\mathrm{n}}\cos \theta.
\label{eq:S3}
\end{equation}
Therefore the component of the particle velocity in the direction of $\mathbf{v}_{\mathrm{n}}$ is
\begin{equation}  v_{\mathrm{p}}\cos\theta + v_{\mathrm{L} \bot \mathrm{n}} =  v_{\mathrm{n}} \cos^2 \theta + v_{\mathrm{L} \bot \mathrm{n}},
\label{eq:S4}
\end{equation}
where  $v_{\mathrm{L} \bot \mathrm{n}}$ is the component of $\mathbf{v}_{\mathrm{L} \bot}$ parallel to $\mathbf{v}_{\mathrm{n}}$.

We take an average over all particles,  assumed to be randomly distributed on the vortex lines.  The average component of the particle velocity in a direction perpendicular to $\mathbf{v}_{\mathrm{n}}$ must vanish,  and therefore we can write for the average particle velocity,  which is in the direction of $\mathbf{v}_{\mathrm{n}}$,
\begin{equation}  \langle v_{\mathrm{p}} \rangle = v_{\mathrm{n}} \langle\cos^2 \theta \rangle + \langle v_{\mathrm{L} \bot \mathrm{n}} \rangle.
\label{eq:S5}
\end{equation}
The average $\langle v_{\mathrm{L} \bot \mathrm{n}} \rangle$ can be evaluated from the results of simulations of the counterflow vortex tangle \cite{AFT}
\begin{equation}  \langle v_{\mathrm{L} \perp \mathrm{n}} \rangle = - \frac{1}{\Omega L v_{\mathrm{n}}} \int \mathbf{v}_{\mathrm{n}} \cdot [\mathbf{s}' \times (\mathbf{s}' \times \dot{\mathbf{s}})]d\xi.
\label{eq:S6}
\end{equation}
Symmetry requires that the vector $\langle [\mathbf{s}'\times (\mathbf{s}' \times \dot{\mathbf{s}})]\rangle$ is parallel to $\mathbf{v}_{\mathrm{n}}$,  so that
\begin{equation}    \langle \mathbf{v}_{\mathrm{L} \bot \mathrm{n}} \rangle = -\frac{1}{\Omega L} \int [\mathbf{s}' \times (\mathbf{s}' \times \dot{\mathbf{s}})]d\xi.
\label{eq:S7}
\end{equation}
Evaluation of the average $\langle \cos^2 \theta \rangle$ requires a knowledge of the anisotropy of the vortex tangle.  In fact it is related  the anisotropy parameter,  $I_{||}$,  introduced by Schwarz and evaluated in a full Biot-Savart simulation by Adachi \textit{et al} \cite{AFT}
\begin{equation}  I_{||}=\frac{1}{\Omega L} \int (1-\cos^2 \theta)d\xi,
\label{eq:S8}
\end{equation}
so that
\begin{equation}  \langle \cos^2 \theta \rangle = 1-I_{||}.
\label{eq:S9}
\end{equation}
Thus we have for the mean particle velocity
\begin{equation}  \langle v_{\mathrm{p}}\rangle = v_{\mathrm{n}}(1-I_{||}) + |\langle \mathbf{v}_{\mathrm{L} \perp \mathrm{n}}\rangle | = \Big(- \frac{\rho_{\mathrm{s}}}{\rho_{\mathrm{n}}}(1-I_{||})  + \frac{|\langle\mathbf{v}_{\mathrm{L}\perp \mathrm{n}}|}{v_{\mathrm{s}}} \Big) v_{\mathrm{s}},
\label{eq:S10}
\end{equation}
where we have used Eq.(\ref{eq:15}).  Eq.(\ref{eq:S10}) is our final result for the mean particle velocity in the limit of small velocities;  The integral $I_{||}$ has already been  calculated in the simulations of Adachi \textit{et al} \cite{AFT},  and we have now calculated the quantity $\langle \mathbf{v}_{\mathrm{L} \perp \mathrm{n}}\rangle$ from the results of the same simulations.

\section{Summary of our results}

\begin{figure}[h]
\begin{center}
\includegraphics[width=1.0\linewidth]{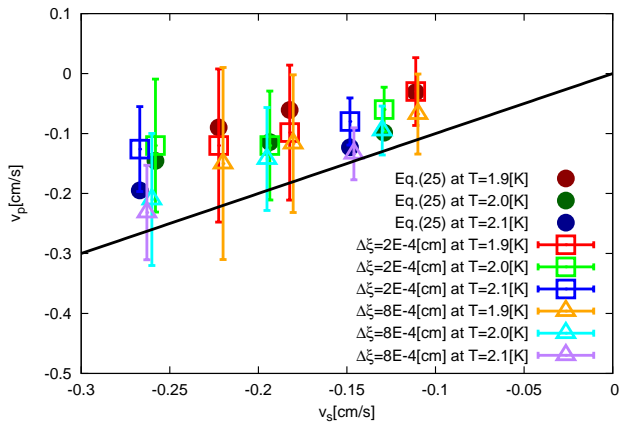}
\caption{(Color on line) The results of our theories and computations.  The solid line shows $v_{\mathrm{p}}=v_{\mathrm{s}}$.}
\label{results}
\end{center}
\end{figure}

The results of our various simulations and calculations are displayed in Fig.\ref{results}.  The points relating to our simulations (Section II) are the particle velocities at the peaks of distributions of the type shown in Fig.\ref{fig:2},  and the error bars are measures of the standard deviations of these distributions.  We have chosen to display them in this way to facilitate comparison with the experimental results of Paoletti \textit{et al.}\cite{Pao},  which are displayed in the same way in their Fig.7(b).  We see that within the large scatter,  seen in both the results of  our simulations and the experimental results,  the different computed velocities and the experimentally observed velocities are in rough agreement.  The velocities are in the direction of,  but significantly less than,  $v_{\mathrm{s}}$,  but are probably proportional to $v_{\mathrm{s}}$.  We judge that this proportionality is a consequence of a simple sliding of the particles along the vortices,  without significant distortion of the vortex tangle.  If this is indeed the case,  then we see that our simulations probably do not extend to velocities at which there is significant distortion;  judging from the experimental results, distortion sets in only at superfluid velocities above roughly   0.4 cm s$^{-1}$.   Although,  as we say,  our results are in rough agreement with experiment,  a careful examination of Fig.\ref{results} suggests that agreement is not as good as we might have hoped.  Our three different sets of theoretical results (relating to two different values of $\Delta\xi$,  and to the calculation of Section III) yield values of the mean particle velocities that seem to differ from one another by small but significant amounts (the standard deviations of the means should be much less that those shown in Fig.\ref{results}). These discrepancies may well be associated with our failure to take into account adequately the precise shape of a vortex in the immediate vicinity of a particle.


\begin{thebibliography}{99}
\bibitem{Hal}$Progress$ $in$ $Low$ $Temperature$ $Physics$, edited by W. P. Halperin and M. Tsubota (Elsevier, amsterdam, 2008), Vol. XVI.
\bibitem{Tsu}M. Tsubota, J. Phys. Soc. Jpn. {\bf 77}, 111006 (2008).
\bibitem{Vin}W. F. Vinen, Proc. R. Soc. London, Ser. A {\bf 240}, 114 (1957); {\bf 240}, 128 (1957); {\bf 242}, 493 (1957); {\bf 243}, 400 (1958).
\bibitem{Lan}L. D. Landau and E. M. Lifshitz, $Fluid$ $Mechanics$ (Pergamon, New York, 1987).
\bibitem{Sch2}K. W. Schwarz, Phys. Rev. B {\bf 38}, 2398 (1988).
\bibitem{AFT} H. Adachi, S. Fujiyama, and M. Tsubota,  Phys. Rev. B {\bf 81}, 104511 (2010).
\bibitem{Zha}T. Zhang and S. W. Van Sciver, Nat. Phys. {\bf 1}, 36 (2005).
\bibitem{Bew}G. P. Bewley, D. P. Lathrop, and K. R. Sreenivasan, Nature (London) {\bf 441}, 588 (2006).
\bibitem{Pao}M. S. Paoletti, R. B. Fiorito, K. R. Sreenivasan, and D. P. Lathrop, J. Phys. Soc. Jpn. {\bf 77}, 111007 (2008).
\bibitem{Guo} W.Guo, S. B. Cahn, J. A. Nikkel, W. F. Vinen and D. N. McKinsey, Phys. Rev. Lett. {\bf 105}, 045301 (2010).
\bibitem{Kiv}D. Kivotides, C. F. Barenghi, and Y. A. Sergeev, Phy. Rev. B {\bf 75}, 212502 (2007).
\bibitem{Sch}K. W. Schwartz, Phys. Rev. A {\bf 10} 2306 (1974).
\bibitem{Fuji}S. Fujiyama, R. H$\ddot{a}$nninen, and M. Tsubota, J. Low. Temp. Phys. {\bf 148}, 263 (2007).
\bibitem{Bar}D. Kivotides, C. F. Barenghi, and Y. A. Sergeev, Phy. Rev. B {\bf 77}, 014527 (2008).
\bibitem{San}S. Barbara, J. Fluid Mech. {\bf 605} 367 (2008).
\bibitem{Kiv1}D. Kivotides, Phy. Rev. B {\bf 77}, 174508 (2008).
\bibitem{Sch1}K. W. Schwarz, Phys. Rev. B {\bf 31}, 5782 (1985).
\end{thebibliography}
\end{document}